\documentclass[times, 11pt, twocolumn]{article}
\usepackage{latex8}
\usepackage{times,url}
\usepackage{graphics,graphicx,setspace}
\usepackage{amssymb,amsmath,amsbsy, amstext, amsthm}

\title{JiTS: Just-in-Time Scheduling for Real-Time Sensor Data 
Dissemination}
\author{Ke Liu, Nael Abu-Ghazaleh and Kyoung-Don Kang\\
Computer Science Dept., SUNY Binghamton \\
\{kliu,nael,kang\}@cs.binghamton.edu}
\date{}

\begin{document}
\maketitle
\begin{abstract} 

We consider the problem of real-time data dissemination in wireless
sensor networks, in which data are associated with deadlines and it is
desired for data to reach the sink(s) by their deadlines. To this end,
existing real-time data dissemination work have developed packet
scheduling schemes that prioritize packets according to their
deadlines.  In this paper, we first demonstrate that not only the
scheduling discipline but also the routing protocol has a significant
impact on the success of real-time sensor data dissemination. We show
that the shortest path routing using the minimum number of hops leads
to considerably better performance than Geographical Forwarding, which
has often been used in existing real-time data dissemination work.  We
also observe that packet prioritization by itself is not enough for
real-time data dissemination, since many high priority packets may
simultaneously contend for network resources, deteriorating the
network performance.  Instead, real-time packets could be judiciously
delayed to avoid severe contention as long as their deadlines can be
met.  Based on this observation, we propose a Just-in-Time Scheduling
(JiTS) algorithm for scheduling data transmissions to alleviate the
shortcomings of the existing solutions.  We explore several policies
for non-uniformly delaying data at different intermediate nodes to
account for the higher expected contention as the packet gets closer
to the sink(s).  By an extensive simulation study, we demonstrate that
JiTS can significantly improve the deadline miss ratio and packet drop
ratio compared to existing approaches in various situations.  Notably,
JiTS improves the performance requiring neither lower layer
support nor synchronization among the sensor nodes.

\end{abstract}

\section{Introduction}\label{intro}

Wireless sensor networks are an important emerging technology that
will revolutionize sensing for a wide range of scientific, military,
industrial and civilian applications.  A large number of inexpensive
sensors collaborating on sensing a phenomena provide cost-effect
detailed monitoring of the area under observation.  While some sensor
networks are deployed to collect information for later analysis, most
applications require monitoring or tracking of phenomena in real-time.  A
primary challenge in such applications is how to carry out sensor data
dissemination given source-to-sink {\em end-to-end} deadlines when the
communication resources are scarce.  The bursty nature of traffic in
sensor networks, as the degree of observed activity varies, can cause
the network resources to be exceeded.  Moreover, the {\em ad hoc}
nature of multi-hop sensor networks makes it difficult to schedule
network traffic centrally as in traditional real-time applications.

Existing solutions for real-time data dissemination~\cite{paper:rap}
prioritize packet transmission at the MAC layer according to the
deadline and distance to the sink.  These work have several
limitations, including: (1) While packets are prioritized, they are
not delayed. When traffic is bursty, high contention may result,
increasing transmission and queuing delays.  Further, packets
generated by different sensors at the same time (e.g., in response to
a detected event), can lead to high collision rates. (2) MAC level
solutions cannot account for the queuing delay in the routing layer
(occurring above the MAC layer) that has a significant impact on
end-to-end delay especially under high load; and (3) MAC level
solutions require re-engineering of the sensor radio hardware and
firmware, making deployment difficult and potentially causing
interoperability problems with hardware supporting
different MAC protocols.  In addition to these effects, the role of
the routing protocol in the real-time scheduling success is not
sufficiently examined. Geographical Forwarding, used
in~\cite{paper:rap,paper:speed,paper:speed2,paper:mmspeed}, does not
always use the shortest path, making it more difficult to meet the
deadline.  Furthermore, using a longer path causes the contention for
the transmission medium to increase as more transmissions are needed
to reach the sink.  (An overview of the related work is presented in
Section~\ref{related}.)


The primary contribution of this paper is a new {\bf Just-in-Time
Scheduling (JiTS)} approach for real-time data dissemination in
sensor networks that addresses many of the shortcomings of the
existing solutions.  JiTS delays packets at every hop for a duration
of time which is a function of distance to the sink and the
deadline.  JiTS uses an estimate of the MAC layer transmission delay
and accounts for it when deciding how long to delay a packet.  By
delaying the packets, rather than only prioritizing them, JiTS
achieves the following advantages: (1) A full estimate of the delay is
used, including the queuing delay at the network layer; (2) The
load is distributed over the available slack time, potentially
allowing the network to tolerate transient periods of high contention
gracefully and to avoid transient hot-spotting; and (3) It provides
packets with a longer time to wait for correlated packets for
aggregation or packet combining.  (In this paper, we do not pursue the
aggregation effect, but reserve it for future work.)

The base JiTS algorithm distributes the slack time (available time
before the deadline expires) uniformly across hops.  However,
in a data collection application, the degree of contention is
typically higher closer to the sink.  Therefore, a second
contribution of this paper is to explore policies that allocate the
time non-uniformly among the hops, to provide more slack time 
for transmission to the hops
closer to the sink; we call this version of the algorithm JiTS with
non-linear delays (JiTS-NL). The third contribution of the
paper is to show that Shortest Path routing outperforms Geographical
Routing with respect to real-time traffic.  
The design of JiTS is discussed in detail in Section~\ref{design}.

In Section~\ref{sim}, the performance of JiTS is compared to
RAP~\cite{paper:rap} and SPEED~\cite{paper:speed} with different
deadline constraints and routing protocols. JiTS outperforms RAP both
in terms of deadline miss ratio and packet drop ratio. Further,
the nonlinear delay version of JiTS is able to achieve the best
performance.  These results hold across different network topologies
(regular and random), and different traffic conditions.  In
particular, when the traffic generation is bursty, JiTS performance
far out-paces that of RAP.  JiTS also performs much better than SPEED,
especially under high load scenarios.  Finally, in
Section~\ref{conclude}, we conclude the paper and discuss 
future work.


\section{Related work}~\label{related}


Lu et al. \cite{paper:rap} propose a real-time communication
architecture in sensor networks, called RAP, which is most
relevant to our work. They propose a
Velocity Monotonic Scheduling (VMS) algorithm to prioritize the data
packets.  VMS derives the required packet ``velocity'', 
which serves as its priority, from the deadline
and distance between source and sink.
In Static VMS, the velocity is computed once at the
source.  Conversely, in Dynamic VMS velocity is recomputed at
intermediate nodes.  Three queues are used for scheduling packets,
where a packet falling in one of the three different priority levels
is queued into the corresponding queue in a FIFO manner, with 
fixed priority enforced among the queues.  It also modifies the
MAC layer back-off scheme to schedule packets according to priority.

The SPEED framework \cite{paper:speed,paper:speed2} proposes an optimized
Geographic Forwarding routing for sensor network. To provide
soft real-time guarantees, SPEED uses a MAC
layer estimate of one-hop transmission delay to select the next hop to
forward the data packet to. However, SPEED does not delay the packets
via scheduling or prioritize data packets.


Li et al. \cite{paper:schmsg} prove that scheduling
parallel messages with deadlines over a wireless channel is NP-hard. 
They propose Last Start Time First (LSTF) scheduling which schedules
messages based on per-hop timeliness constraints by 
manipulating MAC layer back-offs.  They also study the
spatial reuse of the wireless channel and the effect of
collision avoidance.

Multi-hop coordination priority scheduling \cite{paper:dpsch} proposed
to incorporate the distributed priority scheduling into existing
IEEE 802.11 priority back-off schemes to approximate an ideal
schedule. The proposed multi-hop coordination scheduling allows
the downstream nodes to increase a packet's relative priority to make
up for excessive delays incurred upstream. The scheduling requires
modifications of the MAC layer, while possibly overloading the network.

Generally, the ability to meet real-time deadlines in the presence of
contention is related to controlling the load presented to the system.
In terms of networking resources, a related problem is that of congestion
control.  The importance of congestion control in sensor networks was
identified \cite{tilak-02} and approaches for addressing it have
been developed \cite{wan-03}.  Kang et al. study the possibility of
addressing congestion by using multiple paths~\cite{kang-04}.
Exploring the intersection of congestion control and real-time
scheduling is a topic of future research.

SWAN \cite{paper:swan} is a stateless network model differentiating
the service between real-time and best-effort traffic in wireless ad
hoc networks. It supports per-hop and end-to-end control algorithms
without per-flow information. SWAN uses local rate control for UDP and
TCP best-effort traffic and sender-based admission control for
real-time UDP traffic.  Explicit congestion notification is used to
dynamically regulate admitted real-time sessions in the face of
network dynamics due to mobility or traffic overload conditions.  SWAN
and other QoS management schemes in wireless ad hoc networks usually
support unicast network traffic;therefore, they do not map to
many-to-one data dissemination operations prevalent in sensor
networks.


\section{Just-in-Time Scheduling Framework}\label{design}

We consider real-time data dissemination in sensor network
applications where sensor data is being gathered to a sink.  The
effectiveness of the dissemination can be measured in terms of the
deadline miss ratio (indicating how many packets miss their
deadlines) and the packet drop ratio (indicating how many packets are dropped
before arriving at the sink).  The performance is affected by both the
routing protocol and the packet scheduling algorithm.  Consider that
in the absence of contention, the delay of a packet is proportional to
the number of hops on the path from the sensor to the sink, where the
selected path is determined by the routing protocol.  In the presence
of contention, additional delays are incurred as the packets are
queued behind other packets. Also, data transmission can take longer
as the wireless channel is more highly utilized.



In this section, we describe the proposed {\em Just-in-Time}
Scheduling algorithm (JiTS).  We first discuss the role of the routing
protocol and the two routing protocols that we use in
Section~\ref{sec:routing}.  Section~\ref{sec:delay} then presents the
scheduling algorithms of JiTS and compares them to those of RAP.

\subsection{Routing Protocols}
\label{sec:routing}

Existing real-time sensor network protocols, such as the
RAP~\cite{paper:rap} and SPEED~\cite{paper:speed,paper:speed2}
rely on Geographical Forwarding (GF) as the
routing protocol~\cite{paper:gfg,paper:gf,paper:ggr}.  In these
protocol, each node tracks the location of its one hop neighbors via
GPS or some localization algorithm (e.g., \cite{bulusu-00}).  Sensors
know the geographical location of the sink, for example, via the
dissemination of the periodic routing flood from the sink.  Forwarding
is accomplished by sending the data to the neighbor who is closest to
the sink.  The advantage of this approach is small routing overhead;
that is, each node simply needs to track the location information of
its neighbors.  However, this approach may not yield the shortest path
in terms of number of hops, and delivery is not guaranteed. Either RAP 
or SPEED cannot resolve the scheduling for a system based 
on GFG~\cite{paper:gfg} or GPSR~\cite{paper:gf} 
since they are based on the distance progress which
the face forwarding does not provide.  In addition, GPS units are
expensive and energy consuming, while localization algorithms
introduce localization errors that may affect the routing
effectiveness.  Therefore, we investigate a Shortest Path (SP) routing
protocol to illustrate the role that routing plays in determining the
success of a real-time scheduling algorithm.  Like GF, SP operates by
having the sink periodically flood a packet advertising its presence
in the network.  Nodes set up their routing entries as they receive
these advertisement messages, remembering the route with the shortest
path to the sink and the network distance in number of hops.

JiTS requires knowledge of the distance to the sink and the
End-to-end Estimate of the Transmission Delay (EETD) needed by the MAC
layer to transmit a packet.  The first piece of information is
typically present in any traditional routing protocol such as SP or
GF.  The second piece of information (EETD) can also be
collected by the routing layer.  At every hop, we estimate the local
Estimated Transmission Delay (ETD) 
by exchanging a packet infrequently with the next hop neighbor towards
the sink.
A more precise estimate of ETD requires MAC layer support
\cite{paper:speed2}; however, we do not use this approach because it
requires MAC layer changes.  

Summing the ETDs of a data packet hop by hop is costly and may lead 
to inaccurate estimates since one hop ETD can fluctuate significantly.
Therefore, we use the following function to decide the EETD:
\begin{eqnarray*}
EETD=ETD\cdot \frac{E2E~distance}{One~hop~distance}
\end{eqnarray*}
where the distance can be measured in different ways.
Since the queuing delay dominates the end to end delay mostly in a 
heavy traffic environment, a precise EETD is not necessary. 


\subsection{Just-in-Time Scheduling (JiTS)}\label{sec:delay}

Scheduling is the primary mechanism available to intermediate nodes
achieve their deadlines.  Existing approaches to real-time data
dissemination in sensor networks attempt to prioritize packets
according to deadline, but do not intentionally delay packets.  In
contrast, the proposed Just-in-Time Scheduling (JiTS) approach sets
target transmission times for the packets in accordance with their
deadline and the remaining distance along the path. As a result, (1)
JiTS can achieve more effective dissemination by judiciously
allocating the {\em slack time} (time until the deadline) among the
intermediate hops; (2) JiTS can better handle bursty traffic by
decreasing the chances of collisions; and (3) More data aggregation is
possible as discussed before.  In the remainder of this section, we
describe JiTS in a stepwise manner.



\subsubsection{JiTS Organization}

In JiTS, each node must decide how long to delay a packet, while
leaving it with sufficient time to meet the deadline.  The time taken
by a packet from the source to the sink could be divided into two
parts: {\it lower layers transmission delay} spent for
transmission below the network layer and {\it queuing delay}
needed for routing-layer queuing at any intermediate node.  We use the
EETD estimate to approximate the MAC layer transmission delay.  Since
this time is spent below the network layer, the scheduling cannot
directly affect it.  The second component is the queuing delay: how
long the packet is queued in the network layer before it is
handed to the MAC layer for transmission.  This delay can be directly
controlled by the JiTS scheduler.

\begin{figure}[h]
\begin{center}
\includegraphics[width=0.3\textwidth]{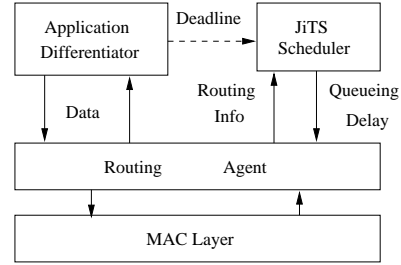}
\end{center}
\caption{JiTS Organization}
\label{fig:arch}
\end{figure}

Figure\ref{fig:arch} shows the organization of JiTS.  We assume that
the deadline information for a packet is present in (or derivable
from) it.  It is possible that the slack time (available time until
the deadline) is smaller than the remaining EETD.  Since JiTS is
mainly designed for soft real-time applications, JiTS immediately 
forwards such packets without any delay even if the packet misses the 
deadline. 

The JiTS scheduler uses a single priority queue for packet forwarding
based on the computed target transmission time.
Transmission is accomplished via a timer that is set to the target
transmission time of the head of the queue.  When the queue is full,
the JiTS scheduler selects the packet at the head of the queue to 
immediately forward it instead of dropping any packet in the queue. 


\subsubsection{Just-in-Time Scheduling Policies}

Different JiTS scheduling policies can be developed based on the ways
of allocating the available slack time among the intermediate nodes.
In the basic JiTS algorithm, the target transmission time is set to 
be equal at all intermediate hops and is determined as follows:
\begin{equation}
\label{equ:jits}
Target~Delay = \dfrac{Deadline - EETD}{Distance(X, sink)}\cdot\alpha
\end{equation}
where $Deadline$ is the end-to-end deadline between a source and sink
and $\alpha$ is a constant "safety" factor used to ensure that
the real-time deadline would be met. When $\alpha = 0.7$, for example, 
the target delay of the packet will be set to 0.7 times the available 
slack time, leaving the remaining time as a safety margin. For
different JiTS policies, node $X$ means differently.

As we can see, the Target Delay of any in-queue packet determines its
priority.  The time a packet is delayed in the queue can be used as
the key to a priority queue that holds the packets to be transmitted.
The end-to-end transmission and processing delay is considered along
with the queuing delay, by taking into account the end-to-end
deadline, distance and EETD.

We consider the following policies for Just-in-Time Scheduling.
\begin{enumerate}
\setlength{\itemsep}{0in}
\item {\bf Static JiTS (JiTS-S):} In JiTS-S, the target delay is
  set at the data source.  In Equation~\ref{equ:jits}, the end-to-end
  deadline is fixed at the data source; the EETD is measured with the
  ETD of forwarding node and the distance from source to sink ($X$ is
  the data source). Thus, even we call it static, the different ETD's
  of forwarding nodes would make the target delay at each node
  different.
  
\item {\bf Dynamic JiTS (JiTS-D):} In JiTS-D, the
  target delay is reset at each forwarding node with the local value
  of parameters. In Equation\ref{equ:jits}, the end-to-end deadline of
  a packet at some forwarding node is the {\em remaining slack time},
  measured by $Deadline - Elapsed~Time$.  The EETD is decided by the
  one-hop ETD of the forwarding node and the distance from it to the
  sink, but not by the distance from the source to sink ($X$ is the 
  current forwarding node). Hence, the dynamic JiTS is able to
  continuously refine the priority of the packet.

\item {\bf Non-linear JiTS (JiTS-NL):} It is also possible to allocate the
  available slack time non-uniformly among the intermediate hops along
  the path to the sink. For example, we may desire to provide the
  packets with additional time as it goes closer to the sink.  The
  intuition is that in a gathering application, the contention is
  higher as the packet moves closer to the sink.  Different policies
  can be developed to break down the available time.  
\end{enumerate}
\noindent
We explore the following policy for JiTS-NL:
\begin{equation}
\label{equ:jits-nl}
Target~Delay = \dfrac{Deadline - EETD}{2^{R/O}}
\cdot \alpha
\end{equation}
where $R$ and $O$ are the remaining distance to the sink and
one hop distance, respectively.  More generally, it is desired to
allocate the slack time proportionately to the degree of contention
along the path. Such a heuristic may be developed by passing the
contention information along with the routing advertisement and
allocating the available slack time accordingly. A thorough
investigation is reserved as future work.






\subsubsection{Just-in-Time Scheduling for Different Routing Protocols}

JiTS can be adapted to work with virtually any underlying routing
protocol.  However, the JiTS algorithm may need to be adapted to
consider the cost metric used by the routing algorithm.  For example,
in a system based on the shortest path routing (SP), the distance
parameters used by JiTS scheduler is measured in number of hops.  The
corresponding functions for basic JiTS (Equation~\ref{equ:jits}) and
non-liner JiTS (Equation~\ref{equ:jits-nl}) are as follows:
\begin{equation}
Target~Delay = \dfrac{Deadline - EETD}{h}\cdot\alpha
\end{equation}
\begin{equation}
Target~Delay = \dfrac{Deadline - EETD}{2^{h}}\cdot \alpha
\end{equation}
where $EETD = ETD \cdot h$ and $h$ stands for the end-to-end 
number of hops. (For the geometric routing, the values of distance 
parameters used in JiTS scheduler is the Euclidean distance.) 

~

In summary, the following information is needed to schedule packets in
JiTS:
\begin{itemize}
\item {\em End-to-end deadline information}: This information is
  provided by the application as part of the data packet to meet
  the requirements of the specific real-time data dissemination 
  application.  
  

\item {\em End-to-end distance information}: this information is
  obtained from the routing protocol. For example, this information is
  maintained in the routing tables of traditional distance vector based
  or link-state based routing protocols to keep track of the cost of
  the path.  Furthermore, in source routed protocols such as 
  DSR\cite{paper:dsr}, this
  information can be directly computed from the packet header which
  includes the full path to the destination.  Finally, in geographic
  routing, Euclidean distance measured as the distance from the
  current node to the destination can be used as the distance metric.
\end{itemize}
\noindent
The output of JiTS scheduler is the queuing delay, which is used by
the routing protocol to decide how long to delay an incoming data
packet before attempting to forward it (by passing it to the MAC
layer).  MAC layer prioritization is not needed by the JiTS design,
since a packet is sent when only its just-in-time local deadline, i.e.,
target delay, is reached. Thus, the MAC layer needs no change to
use our JiTS algorithms.


If the traffic through the current node is not heavy and queuing time
is more than enough for any packet with lenient deadline requirement,
{\em just-in-time} scheduling (or any scheduling) is not needed and
may end up harming performance.  If such a situation can be detected,
JiTS can be disabled and packets forwarded normally.  For example, an
idle detection mechanism may be employed such that if an idle period
passes without packet transmission, the head of the queue is sent
immediately.

\section{Experiments}\label{sim}

We implemented the Static, Dynamic and Non-Linear JiTS with
both Shortest Path (SP) routing and Geographic Forwarding (GF) in
the Network Simulator (NS2, version 2.27)~\cite{web:ns2}.  We also
implemented the RAP Velocity Monotonic Scheduling (VMS) with GF,
including the specialized MAC support following
the specification of the authors~\cite{paper:rap}, as well as the SPEED
protocol~\cite{paper:speed,paper:speed2}.  Since GF has been
shown to significantly outperform traditional routing protocols, such
as DSR~\cite{paper:dsr}, in sensor network data
dissemination, we restrict the routing comparison to GF and SP.

\subsection{Comparisons with VMS}
Table~\ref{tab:scenario} shows the simulation parameters we use to 
compare JiTS to Velocity Monotonic Scheduling (VMS) adopted by 
RAP~\cite{paper:rap}. Unless otherwise indicated, these parameters 
are used in this studies.
\begin{table}[h]
\centering 
\begin{tabular}{|c|c|}
\hline MAC protocol & IEEE 802.11\\
\hline Transmission Range & 250 m\\
\hline Bandwidth & 2Mbps\\
\hline Data Packet Size & 32B\\
\hline Data Rate & 2 packet/s\\
\hline Routing Period & 5 s\\
\hline Simulation Area & 1000 $\times$ 1000 $m^2$\\
\hline Sensor Nodes & 100\\
\hline Simulation time & 120 s\\
\hline
\end{tabular}
\caption{Simulation parameters}
\label{tab:scenario}
\end{table}
We use scenarios with 100 sensors and investigate both $10\times10$
grid and random deployment.  In the grid scenarios, the sink is placed
on the northwest corner of the network.  In random deployment, the 100
nodes are randomly placed in the simulation area while the sink is
placed roughly at the center of the area.



We compared JiTS scheduling with the VMS both using the same routing
protocol (GF) that was used in the original RAP
scheme~\cite{paper:rap}.  Later, we also show that SP significantly
outperforms GF for JiTS. As discussed before, we use soft-deadline
policy for all protocols where packets are not dropped if their
deadline is exceeded.  We note that a hard deadline version of JiTS
could be easily developed; however, we conjecture that soft-deadlines
are likely to be a more realistic model for sensor networks.  Since
JiTS does not require any MAC layer information, we use the original
IEEE 802.11 as our MAC layer protocol, while we use the modified MAC
layer for RAP as done in~\cite{paper:rap}.  We have implemented
Static, Dynamic and Nonlinear JiTS policies and compared their
performance with RAP.


We considered the issue of what the JiTS safety margin parameter
$\alpha$ should be set to.  If $\alpha$ is too high, most of the slack
time is taken up by the intentional JiTS delay. As a result,
additional unexpected delays, if any, may cause a packet deadline
miss.  Conversely, if the $\alpha$ is too low, packets are
conservatively sent quickly towards the sink, possibly overflowing
buffers around it.  Experimentally, we observed that a safety margin
parameter of 0.7 works well across different deadlines.  Thus, 30\% of
the slack time is set aside to account for unexpected
transmission or queuing delays. (An analytic derivation of $\alpha$
is reserved for future work.)

\subsubsection{Evaluation of JiTS and VMS}

\begin{figure}[t]
\centering
\includegraphics[width=0.35\textwidth]{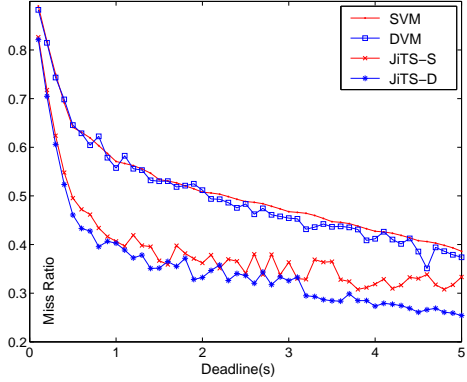}
\caption{JiTS(GF) vs VMS Miss Ratio}
\label{fig:JitsVsRapMr}
\end{figure}
\begin{figure}[t]
\centering
\includegraphics[width=0.35\textwidth]{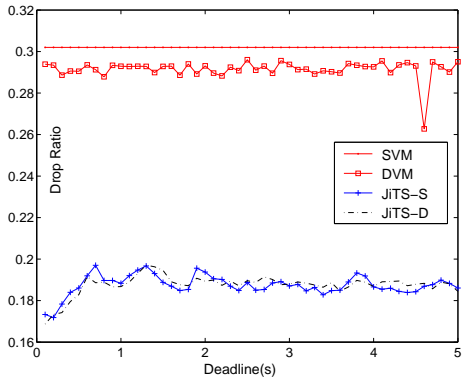}
\caption{ Drop Ratio: JiTS(GF) vs VMS}
\label{fig:JitsVsRapDr}
\end{figure}

The first experiment compares the performance of JiTS to RAP.
Figure~\ref{fig:JitsVsRapMr} and figure~\ref{fig:JitsVsRapDr} show
that for different deadline requirements, the miss ratios of Static
and Dynamic JiTS are much lower than those of Dynamic VMS(DVM) 
and Static VMS(SVM) across all
the tested deadlines.  The same observation holds for the drop
ratios. Dynamic JiTS outperforms static JiTS in terms of the miss
ratio.

\begin{figure}[t]
\centering
\includegraphics[width=0.35\textwidth]{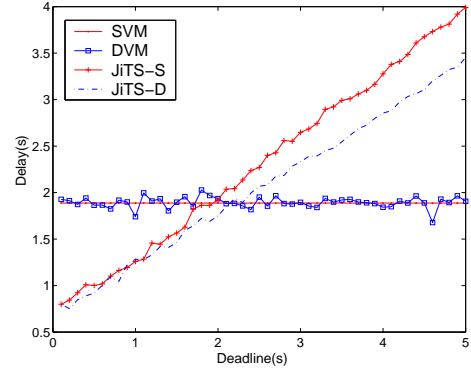}
\caption{Average Delay: JiTS (GF) vs VMS} 
\label{fig:JitsVsRapAvgDelay}
\end{figure}

Figure~\ref{fig:JitsVsRapAvgDelay} shows the average delay of JiTS and
RAP to illustrate the difference between the two scheduling
approaches.  The average delay of JiTS grows linearly with the
deadline as the intermediate nodes delay packets proportionately to
the deadline. In this way, we can take advantage of the slack under
overload. Note that dynamic JiTS manages to keep the average delay 
around the value of $0.7 \cdot deadline$, while the static JiTS has
slightly higher average delay.  Since RAP does not delay packets,
its delay only depends on the data generation pattern and does not
change with the deadline.  

Since the VMS uses multiple FIFO queues as its priority queue, packet
starvation commonly happens and the maximum packet delay suffered by
RAP is much worse than that of JiTS by a factor of 2 to 3. (Due to
space limitations, we do not include the results here.)  We will only
compare our JiTS policies with SVM in the remainder of this paper,
because the performance of SVM and DVM is similar and the
authors of RAP observed SVM to be superior to DVM~\cite{paper:rap}.

\subsubsection{Effect of Routing Protocol}

\begin{figure}[t]
\centering
\includegraphics[width=0.35\textwidth]{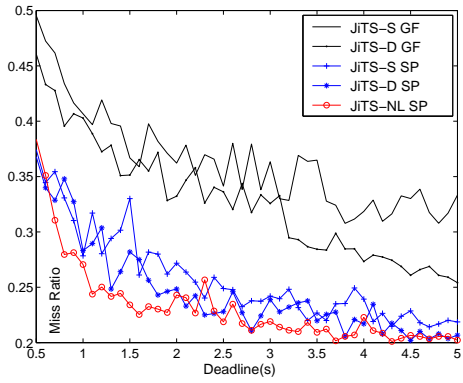}
\caption{JiTS SP vs GF Miss Ratio}
\label{fig:JitsSpvsGfMr}
\end{figure}
\begin{figure}[t]
\centering
\includegraphics[width=0.35\textwidth]{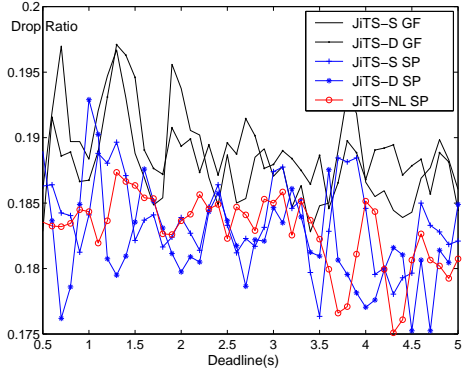}
\caption{Drop Ratio: JiTS SP vs GF} 
\label{fig:JitsSpvsGfDr}
\end{figure}

In the second set of experiments, we compare the performance of JiTS under GF
with JiTS with Shortest Path (SP) routing.  In addition, in this
experiment, we include the performance of the nonlinear JiTS
scheduling algorithm (JiTS-NL).  Figure~\ref{fig:JitsSpvsGfMr} and
Figure~\ref{fig:JitsSpvsGfDr} show the miss ratio and drop ratio
respectively.  From these figures, we observe that 
JiTS performs considerably better with SP than
with GF. In general, dynamic JiTS performs better than static JiTS for
both routing protocols.  Furthermore, JiTS-NL provides significant
improvements compared to static and dynamic JiTS. 
Especially, the improvement is most pronounced under tight deadlines
showing the applicability of JiTS-NL to real-time data dissemination.  
The maximum and average delay of
JiTS with GF is higher than that with SP for the same deadlines (results
not shown) because GF may use longer paths than SP in terms of the 
number of hops.  

\subsubsection{Performance under Bursty Traffic}
\begin{figure}[t]
\centering
\includegraphics[width=0.35\textwidth]{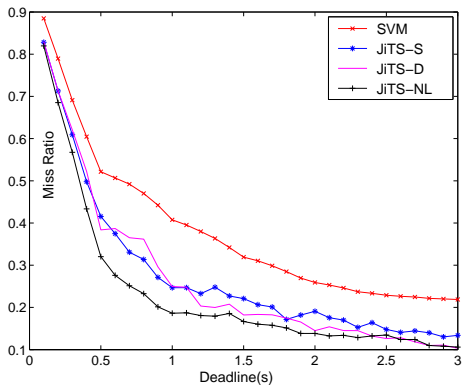}
\caption{Miss Ratio of Bursty: JiTS vs SVM}
\label{fig:bursty}
\end{figure}

\begin{figure}[t]
\centering
\includegraphics[width=0.35\textwidth]{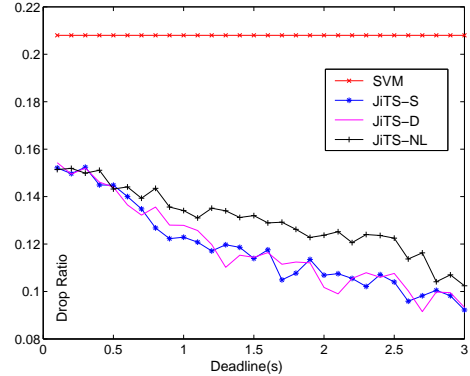}
\caption{Drop Ratio of Bursty: JiTS vs SVM}
\label{fig:burstyDelay}
\end{figure}

In this set of experiments, we evaluate the performance of JiTS
vs. RAP under bursty traffic conditions. At every 10 seconds we let
all the nodes publish data packets with the pre-set data rate in the
first 5 seconds then stop publishing for the remaining 5 seconds.
Figures~\ref{fig:bursty} and~\ref{fig:burstyDelay} show the
miss ratios and drop ratios of JiTS and SVM under bursty traffic with
end-to-end deadline increasing from 0.1 second to 3.0 seconds.  In
Figure~\ref{fig:bursty}, we can see that the miss ratio of dynamic
JiTS is much lower than that of SVM under the bursty traffic, because
JiTS can tolerate the traffic burst by delaying some packets, and
taking advantage of the idle period.  On the other hand, SVM cannot
make use of the traffic behavior, since it does not delay packets even
if there is slack. In Figure~\ref{fig:burstyDelay}, we observe that
JiTS disciplines achieve the lower the drop ratios than SVM,
delivering more packets as the deadline constraints are relaxed.
In contrast, SVM suffers almost the same drop ratio even
when the deadlines are relaxed.


\subsubsection{Performance under Random Deployment}

\begin{figure}[t]
\centering
\includegraphics[width=0.35\textwidth]{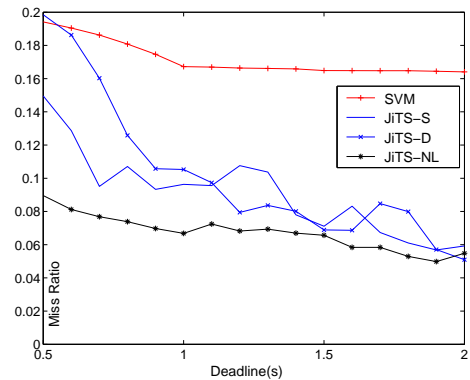}
\caption{Random Scen. Miss Ratio (GF)}
\label{fig:randomMR}
\end{figure}
\begin{figure}[t]
\centering
\includegraphics[width=0.35\textwidth]{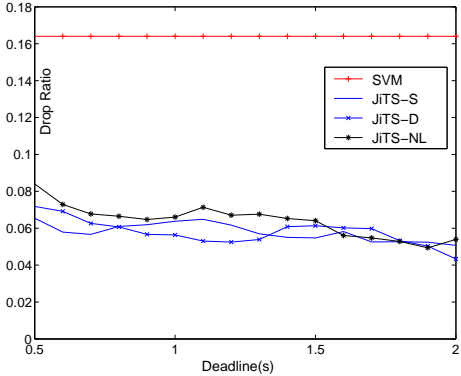}
\caption{Random Scen. Drop Ratio (GF)}
\label{fig:randomDR}
\end{figure}
JiTS and SVM were also evaluated using constant traffic 
(Table~\ref{tab:scenario}) and 
a random deployment scenario where the 100 nodes were randomly 
placed within the simulation area. To be fair, we associate 
JiTS with GF routing which SVM is based on. 
We varied the deadline requirements 
from 0.5 to 2.0 seconds in steps of 0.1 seconds. 
Figures~\ref{fig:randomMR} and~\ref{fig:randomDR} show the miss
ratios and drop ratios of the tested algorithms.  The simulations
show that both JiTS and SVM perform much better in a random scenarios
than they did in the grid scenarios possibly because the location of
the sink is central to the simulation area, making the average sensor
distance to the sink smaller.  Again JiTS provides superior
performance to SVM.  For the SVM, the drop ratios do not
decrease as the deadline grows since it prioritizes but does not delay
packets as shown in Figure~\ref{fig:randomDR}.  
The drop ratio becomes the lower bound of the miss ratio.
JiTS shows more reactivity in that both the drop ratio and miss ratio
decrease as the deadline requirement is relaxed.

\subsubsection{Performance with Multiple Deadline Data}

\begin{figure}[t]
\centering
\includegraphics[width=0.35\textwidth]{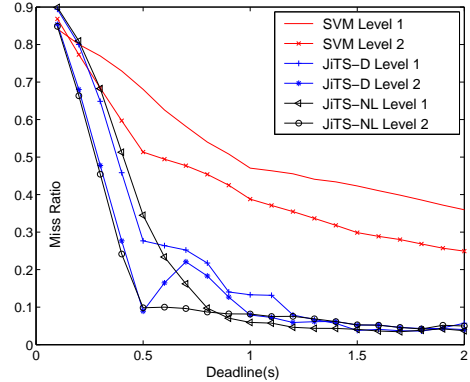}
\caption{Two Level Deadlines: Miss Ratio}
\label{fig:twolevelMR}
\end{figure}

\begin{figure}[t]
\centering
\includegraphics[width=0.35\textwidth]{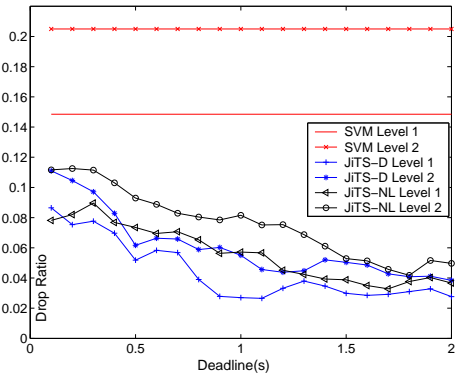}
\caption{Two Level Deadlines: Drop Ratio}
\label{fig:twolevelDR}
\end{figure}

We simulated the presence of two data types being generated by the
sensors with two different deadline constraints (deadline of level 1
data is half that of level 2 data), in bursts.  Ideally, the
scheduling algorithm would allow both data types to meet their
deadlines effectively.  Figures~\ref{fig:twolevelMR} 
and~\ref{fig:twolevelDR} show the miss ratio and drop ratio of SVM,
JiTS Dynamic and Non-Linear.  Under strict deadlines, level 2 traffic
receives better performance, because the network was often unable to
satisfy the aggressive level 1 deadlines.  Once the deadlines increase
beyond a certain level, JiTS is able to provide similar performance to
the two traffic types, despite the different deadline requirements.
However, the same is not true of SVM, where level 2 traffic continues to
receive better performance than level 1 traffic.


\subsection{Comparisons with SPEED}

We also built simulation models for the SPEED
framework~\cite{paper:speed,paper:speed2} within the NS-2 simulator.
We implemented the full specification of SPEED, SPEED-T (Minimal one
hop delay first), and SPEED-S (maximal one hop progress speed first).
For validation, we first repeated the experiments conducted in the
original SPEED paper \cite{paper:speed}.  We note that the original
SPEED implementation was built in a different simulator
(GloMoSim~\cite{glomosim}) which has different wireless models than
NS-2.  Further, the authors did not specify exact values of the
several (at least four) simulation parameters that they used in their
experiments. (In a personal communication with one of the authors, he
indicated that the parameters were manually tuned for each case and we
were not able to repeat the exact configuration.)  Nevertheless, our
obtained results were very close to those achieved by the original paper
(but not identical).  In this section, we first compare the
performance of SPEED with GF and SP and then compare
SPEED with JiTS under different traffic load levels.

\subsubsection{Evaluation of SPEED and routing protocols}

We compared the simulation results against the shortest-path
routing and geographical forwarding routing in a void-free deployment.
In this study, all the data flows are constant, while 
some congestion-introducing 
flows are created in some intermediate nodes in the same manner used in 
the original SPEED design\cite{paper:speed,paper:speed2}.

\begin{figure}[t]
\centering
\includegraphics[width=0.4\textwidth]{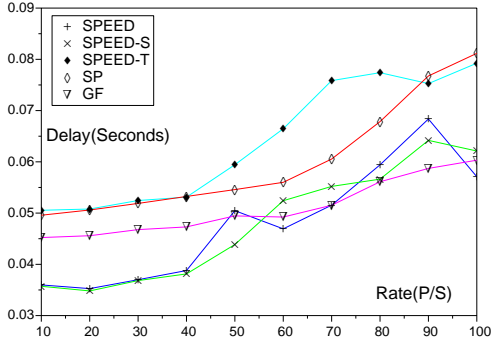}
\caption{SPEED: end-to-end delay}
\label{fig:speed-us-delay}
\end{figure}

\begin{figure}[t]
  \centering
  \includegraphics[width=0.38\textwidth]{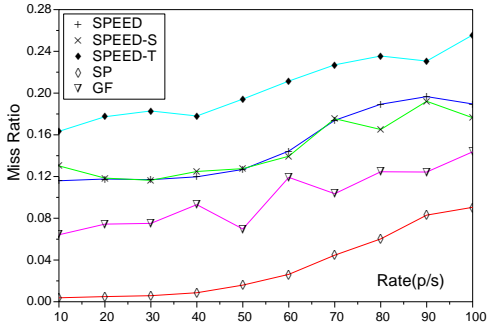}
\caption{SPEED: miss ratio}
\label{fig:speed-us-missratio}
\end{figure}

Figure \ref{fig:speed-us-delay} and Figure
\ref{fig:speed-us-missratio} show the performance of SPEED.
For the congestion-introducing flow with lower data rate, SPEED and GF
perform similarly as observed in the original SPEED
paper\cite{paper:speed}.  
In terms of end-to-end delay, SPEED is better than both the GF and SP
in most cases, especially when the data rate of congestion-introducing
flow is low.  However, the miss ratio of SPEED becomes higher than
GF and SP in all cases.  It is very high even when the data rate
of congestion-introducing flow is low. The miss ratios of GF and SP
are not as high as the values reported in \cite{paper:speed}.  The
possible reason of it is that the SPEED tries to drop some data
packets when ``SPEED'' can not be maintained in some intermediate
nodes, which would backpressure the data source. The future dropping
possibility would be increased as this situation keeps happening.
Meanwhile, neither GF or SP in our simulation drop data packets (in
the routing layer), leading to a lower drop ratio. 

\begin{figure}[t]
\centering
\includegraphics[width=0.4\textwidth]{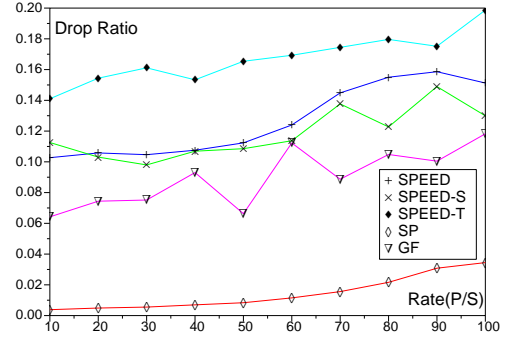}
\caption{SPEED: drop ratio}
\label{fig:speed-us-dropratio}
\end{figure}

Figure \ref{fig:speed-us-dropratio} shows the drop ratio of SPEED
against both the GF and SP.  The drop ratio of SPEED is much higher
than that of SP. It is also a little bit higher than GF.  As we can
see, the high miss ratio mostly results from SPEED's high drop ratio.
As expected, SP outperforms GF.  Since the drop ratio of SP is much
lower than the GF, it seems that more data packets with longer
end-to-end delay are received by SP, which leads to the higher average
end-to-end delay as shown in Figure\ref{fig:speed-us-delay}.

\begin{figure}[t]
\centering
\includegraphics[width=0.4\textwidth]{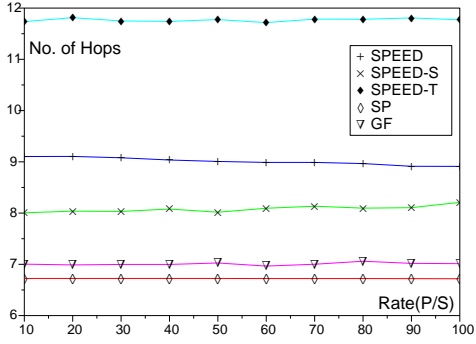}
\caption{SPEED: average hops}
\label{fig:speed-us-avghops}
\end{figure}

Figure~\ref{fig:speed-us-avghops} shows the average number of hops from
source to sink. The average hops of SP is the lowest in all, since it
provides the shortest path at any time. GF also uses shorter paths than 
SPEED. This is mainly because SPEED does not always try to minimize the
number of hops or select the geographically closest neighbor.

\subsubsection{Evaluation of JiTS and SPEED}

In this subsection, we SPEED with JiTS under different traffic load
levels.  Specifically, we use three different traffic load levels in a
10 by 10 grid deployment.  Load I and Load II have 100 sources
generating packets at 1 packet/sec and 0.5 packet/sec respectively.
In case Load III, there are 10 sources generating packets at 1
packet/sec.

\begin{figure}[t]
\centering
\includegraphics[width=0.35\textwidth]{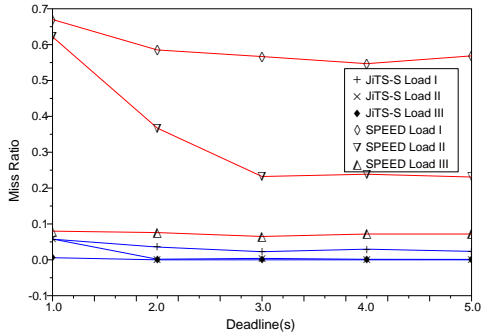}
\caption{JiTS vs SPEED: Miss Ratio}
\label{fig:jits-speed-mr}
\end{figure}

\begin{figure}[t]
\centering
\includegraphics[width=0.35\textwidth]{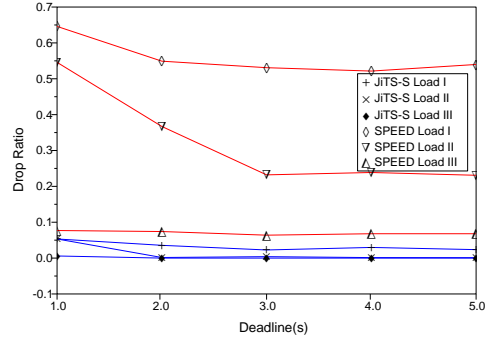}
\caption{JiTS vs SPEED: Drop Ratio}
\label{fig:jits-speed-dr}
\end{figure}

Figure~\ref{fig:jits-speed-mr} and figure~\ref{fig:jits-speed-dr} show
their miss ratios and drop ratios of JiTS and SPEED under the three
different traffic levels.  Under medium and high traffic load, the
drop ratios and miss ratios of SPEED are very high as SPEED attempts
unsuccessfully to route around congestion.  In contrast, JiTS adapts
with different levels of traffic loads.  In light traffic, the two
approaches perform comparably.  


\begin{figure}[t]
\centering
\includegraphics[width=0.35\textwidth]{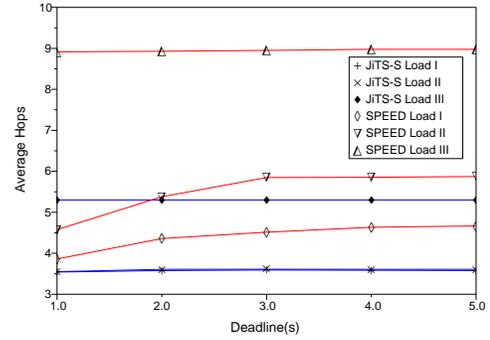}
\caption{JiTS vs SPEED: Average Hops}
\label{fig:jits-speed-ah}
\end{figure}

Figure~\ref{fig:jits-speed-ah} shows the average number of end-to-end
hops of all received data packets. Since JiTS uses SP as basic
routing, the average number of hops is kept stable.  SPEED has a
higher average number of hops due to the backpressure mechanism. If
the timing constraint becomes more stringent, SPEED seems to drop more
packets at intermediate nodes, statistically reducing the average
number of hops traveled by the delivered packets.  


\section{Conclusions and Future Work}~\label{conclude}

Real-time data dissemination is a service of great interest to many
sensor network applications.  We proposed and evaluated the {\em
  Just-in-Time} scheduling mechanisms for the real-time sensor networks
applications that offers significant advantages over existing
real-time sensor data dissemination schemes.  JiTS accomplishes
real-time support by delaying packets a fraction of their slack time
at each hop.  As a result, it is better able to tolerate bursts than
schemes that simply prioritize packet transmission.  We also explored
the effect of routing on real-time scheduling success and showed that
Geographical Forwarding can lead to suboptimal operation. JiTS 
outperforms RAP in both the miss ratio and overall delay.  
We explored several criteria
for allocating the available slack time among the different nodes and
showed that nonlinear distribution of the slack time, assigning more time
assessed to hops closer to the sink, results in better performance than
linear distribution of the slack time. Further, JiTS is a routing layer
solution and does not require changes to lower level protocols.
As a result, it can be deployed independent of the underlying 
sensor network hardware capabilities.
From the simulations, we found the drop ratio is the lower bound of
the miss ratio of real-time communication. If the drop ratio is
decreased, given a reasonable end-to-end deadline, the miss ratio of
these real-time applications should also be decreased.  Mostly the
packets are dropped due to congestion as the network capacity is
exceeded.  In the future, we will further investigate JiTS in the
context of wireless sensor networks. We will also investigate other
related issues such as data aggregation in JiTS and congestion control
for real-time data transmission in sensor networks.




\bibliographystyle{unsrt}
\bibliography{jitsbib.bib}

\begin{thebibliography}{10}

\bibitem{paper:rap}
Chenyang Lu, Brian~M. Blum, Tarek~F. Abdelzaher, John~A. Stankovic, and Tian
  He.
\newblock {RAP}: A real-time communication architecture for large-scale
  wireless sensor networks.
\newblock {\em RTAS'02}, 2002.

\bibitem{paper:speed}
Tian He, John~A Stankovic, Chenyang Lu, and Tarek Abdelzaher.
\newblock {SPEED}: A stateless protocol for real-time communication in sensor
  networks.
\newblock {\em ICDCS'03}, 2003.

\bibitem{paper:speed2}
Tian He, John~A. Stankovic, Chenyang Lu, and Tarek~F. Abdelzaher.
\newblock A spatiotemporal protocol for wireless sensor network.
\newblock {\em IEEE Transactions on Parallel and Distributed Systems}, 2005.

\bibitem{paper:mmspeed}
Emad Felemban, Chang-Gun Lee, Eylem Ekici, Ryan Boder, and Serdar Vural.
\newblock Probabilistic {QoS} guarantee in reliability and timeliness domains
  in wireless sensor networks.
\newblock {\em IEEE INFOCOM'05}, 2005.

\bibitem{paper:schmsg}
Huan Li, Prashant Shenoy, and Krithi Ramamritham.
\newblock Scheduling messages with deadlines in multi-hop real-time sensor
  networks.
\newblock {\em UMASS CMPSCI Technical Report TR04-91}, 2004.

\bibitem{paper:dpsch}
V.~Kanodia, C.~Li, A.~Sabharwal, B.~Sadeghi, and E.~Knightly.
\newblock Distributed multi-hop scheduling and medium access with delay and
  throughput constraints.
\newblock {\em MobiCom'01}, 2001.

\bibitem{tilak-02}
S.~Tilak, N.~Abu-Ghazaleh, and W.~Heinzleman.
\newblock Infrastructure tradeoffs in sensor networks.
\newblock In {\em WSNA'02}, 2002.
\newblock Held in Conjunction with MobiCom 2002.

\bibitem{wan-03}
Chieh-Yih Wan, Shane~B. Eisenman, and Andrew~T. Campbell.
\newblock Coda: Congestion detection and avoidance in sensor networks.
\newblock In {\em SenSys'03}, 2003.

\bibitem{kang-04}
{JaeWon Kang and Yanyong Zhang and Badri Nath}.
\newblock {Adaptive Resource Control Scheme to Alleviate Congestion in Sensor
  Networks}.
\newblock In {\em Proc. of the First Workshop on Broadband Advanced Sensor
  Networks}, 2004.

\bibitem{paper:swan}
Gahng-Seop Ahn, Andrew~T. Campbell, Andras Veres, and Li-Hsiang Sun.
\newblock Supporting service differentiation for real-time and best effort
  traffic in stateless wireless ad hoc networks ({SWAN}).
\newblock {\em IEEE Transactions on Mobile Computing}, 2002.

\bibitem{paper:gfg}
Prosenjit Bose, Pat Morin, Ivan Stojmenovic, and Jorge Urrutia.
\newblock Routing with guaranteed delivery in ad hoc wireless networks.
\newblock In {\em 3rd ACM Int. Workshop on Discrete Algorithms and Methods for
  Mobile Computing and Communications DIAL M99}, Seattle, WA, August 1999.

\bibitem{paper:gf}
B.~Karp and Kung.~H. T.
\newblock {GPSR}: Greedy perimeter stateless routing for wireless networks.
\newblock {\em MobiCom'00}, 2000.

\bibitem{paper:ggr}
Guoliang Xing, Chenyang Lu, Robert Pless, and Qingfeng Huang.
\newblock On greedy geographic routing algorithms in sensing-covered networks.
\newblock {\em MobiHoc'04}, 2004.

\bibitem{bulusu-00}
N.~Bulusu, J.~Heidemann, and D.~Estrin.
\newblock {GPS}-less low cost outdoor localization for very small devices.
\newblock {\em {IEEE} Personal Communications Magazine}, pages 28--34, October
  2000.

\bibitem{paper:dsr}
David~B. Johnson, David~A. Maltz, and Josh Broch.
\newblock {DSR}: The dynamic source routing protocol for multi-hop wireless ad
  hoc networks.
\newblock In Charles~E. Perkins, editor, {\em Ad Hoc Networking}, pages
  139--172. Addison-Wesley, 2001.

\bibitem{web:ns2}
The {N}etwork {S}imulator - ns-2.
\newblock http://www.isi.edu/nsnam/ns/, 2005.

\bibitem{glomosim}
{GloMoSim}.
\newblock http://pcl.cs.ucla.edu/projects/glomosim/.

\end{thebibliography}

\end{document}